\documentclass[epjST]{svjour}
\usepackage{graphicx}
\usepackage{amsmath}
\usepackage{mathtools}
\usepackage{amssymb}
\usepackage{bm}
\usepackage{pifont}
\usepackage{color}
\usepackage[FIGTOPCAP]{subfigure}
\usepackage[latin1]{inputenc}

\newcommand{\ve}{\epsilon}
\newcommand{\vre}{\varepsilon}

\newcommand{\omi}{\mbox{\boldmath $\Omega_i$}}

\newcommand{\HT}{\mbox{\boldmath $H$}}
\newcommand{\QTa}{\mbox{\boldmath $Q_1$}}
\newcommand{\QTb}{\mbox{\boldmath $Q_2$}}

\begin{document}

\title{Towards a quantitative kinetic theory of polar active matter}

\author{Thomas Ihle\inst{1}\inst{2}\fnmsep\thanks{\email{thomas.ihle@ndsu.edu}}}
\institute{Department of Physics, North Dakota State University, Fargo, ND 58108-6050, USA \and
Max-Planck-Institute for the Physics of Complex Systems, N{\"o}thnitzer Stra{\ss}e 38, 01187 Dresden, Germany}

\abstract{
A recent kinetic approach for Vicsek-like models of active particles is reviewed.
The theory is based on an exact Chapman-Kolmo\-gorov equation in phase space. It can handle 
discrete time dynamics and ``exotic'' multi-particle interactions.
A nonlocal mean-field theory for the one-particle distribution function 
is obtained by assuming molecular chaos.
The Boltzmann approach of Bertin {\em et al.}, Phys. Rev. E {\bf 74}, 022101 (2006) and
J. Phys. A {\bf 42}, 445001 (2009),
is critically assessed and compared to the current approach.
In Boltzmann theory, a collision starts when two particles enter each others action spheres and is finished
when their distance exceeds the interaction radius. 
The average duration of such a collision, $\tau_0$,  is measured for the Vicsek model with continuous time-evolution. 
If the noise is chosen to be close to the flocking threshold, the average time between collisions %
is found to be roughly equal to $\tau_0$ at low densities.
Thus, 
the continuous-time Vicsek-model
near the flocking threshold 
cannot be accurately described by a Boltzmann equation, 
even at very small density 
because 
collisions take so long that typically
other particles join in, 
rendering Boltzmann's binary collision assumption invalid.
Hydrodynamic equations for the phase space approach are derived by means of a Chapman-Enskog expansion.
The equations are compared to the Toner-Tu theory of polar active matter. New terms, absent in the Toner-Tu theory, are highlighted.
Convergence problems of Chapman-Enskog and similar gradient expansions are discussed.
}

\maketitle

\section{Introduction}
\label{intro}

The past decade has seen a surging interest in active matter, 
\cite{ramaswamy_10,vicsek_zafeiris_12,marchetti_13}. 
According to Ref. \cite{marchetti_13}, active matter systems are defined by ``their
unifying characteristic that they are
composed of self-driven units, active particles, each capable of converting stored or ambient free energy into systematic 
movement''.
Describing these intrinsic nonequilibrium systems analytically poses a big challenge
because there is no powerful general framework such as the free-energy formalism for equilibrium systems.
Systems have to be
treated on a case by case basis, and often uncontrolled approximations are employed.
Even the simplest active matter models show a wealth of interesting phenomena and are not completely understood.
For this reason I focus on one of the simplest models of active matter -- the Vicsek
model (VM) \cite{vicsek_95_97,nagy_07} -- with the goal of custom-making a quantitative theory for it.
The rationale behind this strategy is that by not modifying a computationally-efficient microscopic model, 
the theory can be directly compared to existing agent-based
numerical simulations and new simulation data can be easily created. Contrary to theories without a direct link to experiment or simulation, 
if we find
there is no good quantitative agreement, we immediately know that the approximations must be 
too crude or the calculations must be 
flawed.
I believe that such a direct feeback will be very helpful in constructing accurate theories of active matter.
For example, the kinetic approach presented in this article has been shown to quantitatively 
reproduce the shape and properties of strongly
nonlinear waves in the Vicsek-model, in the limit of large mean free path \cite{ihle_13}.

In 1998 Toner and Tu proposed hydrodynamic equations for the density and momentum density of
active particles and performed a dynamic renormalization \cite{toner_98,toner_12}. 
While they were able to explain how long-range orientational order
can be achieved in a two-dimensional system of active particles, 
this theory, to my knowledge, has not been able to reproduce the details of
the phase transition
from a disordered state to a state of collective motion in the VM \cite{foot5}.

The terms in the hydrodynamic equations of Ref. \cite{toner_98} were postulated based on 
rotational symmetry and relevance in the renormalization group sense. 
The coefficients of these terms are undetermined by construction. In reality, the coefficients are all related to
just a few parameters of the underlying microscopic interactions, and therefore typically cannot be modified independently of each other.
A direct derivation of the macroscopic equations from the microscopic model is benefitial because it
strongly reduces the parameter space of allowed cofficients and can reveal relevant terms that might have been overlooked. 

Using the Vicsek-model as a paradigm of active matter, a number of important fundamental questions can be studied,
such as (i) is the Toner-Tu hydrodynamic theory sufficient to quantitatively describe active matter,
is there relevant terms missing, does one need more than two equations or nonlocal equations instead?, (ii) 
Is it possible to rigorously derive the coefficients in the Toner-Tu theory from a microscopic model?, 
(iii) how to calculate 
the scaling behavior near
the transition to collective motion, and (iv) how can one extend the theory for the Vicsek-model to more realistic models?

The first two questions are actually the active matter equivalent of Hilbert's famous 6. problem about how
to establish a direct link between microscopic dynamics and macroscopic equations \cite{hilbert_02,gorban_13,slemrod_13}. 
To my knowledge, there has been quite 
some progress
on the solution of this problem for regular Hamiltonian systems \cite{gorban_13} but, recently,
serious doubts on its 
solution have been raised \cite{slemrod_13}. %
By restricting ourselves to simple microscopic models like the VM which has particles of zero 
volume and employs external uncorrelated noise terms, it is interesting to explore whether 
Hilbert's problem is solvable for active systems, at least in certain limits.

As a step in this direction, in this paper I will discuss a kinetic theory approach for self-driven particles \cite{ihle_11}.
The theory does not start at the coarse-grained Boltzmann-level for the one particle distribution function, $f({\bf x}, {\bf v},t)$.
Instead it is based on an exact equation, the Chapman-Kolmogorov equation, in the full phase space
of the model. A mean-field kinetic equation is then derived by using Boltzmann's principle
of molecular chaos. %
The kinetic equation has been evaluated analytically and numerically \cite{ihle_11,ihle_13} and extended
to topological interactions \cite{chou_12}.
Hydrodynamic
equations are derived from it by means of the Chapman-Enskog expansion and will be discussed in this paper.
Surprisingly, these equations contain terms which are not included in the Toner-Tu theory, even if one makes
the coefficients of this theory density-dependent. 

The kinetic theory presented in this paper, which I will call phase space approach (PSA), 
has been thoroughly tested for a model
with passive particles \cite{malevanets_99,gompper_09}. 
The regular Navier-Stokes equations have been derived \cite{ihle_09} and all transport coefficients were found
to agree within a few percent with direct simulations and with alternative theoretical predictions \cite{pool_05,ihle_03_05}.
In addition to the PSA approach, there has been other attempts to derive the Toner-Tu equations from a
microscopic model. One of the first attempts is due to Bertin {\em et al.} \cite{bertin_06,bertin_09}, one of the most 
recent approaches was presented by Gro{\ss}mann {\em et al.} \cite{grossmann_13}.
The common characteristic of these approaches is that they do not treat the original VM with discrete time step and 
genuine multiparticle collisions but other, often simpler, models related to it.
The method presented in this paper deals with the Vicsek model as is. It keeps the original
time-discrete dynamics and the multi-particle collisions. Moreover, it distinguishes between 
the so-called forward- and backward updating schemes and could 
be used to calculate the differences in the phase diagrams and density waves due to different updating methods,
\cite{baglietto_08_09,aldana_09}.

Kinetic theory approaches work best if there is a strong mixing of particles, e.g. when collision partners change rapidly.
When cluster formation occurs, particles have a stronger tendency to recollide and to stay together for longer.
Theoretical approaches for strong clustering of self-propelled particles which is in some sense the opposite limit to what 
is treated by kinetic theories,
have been presented recently by Peruani {\em et al.} \cite{peruani_10,peruani_13}.
The ultimate theory for active particles should contain both scenarios -- clustering and strong mixing -- as limit cases.
Exploring the mixing side of this problem, the PSA approach presented in this paper can hopefully
contribute to the construction of such a general framework.

\section{Vicsek model}
\label{sec:Vicsek}

\subsection{Definition}
\label{subsec:modeldef}

Consider the  two-dimensional Vicsek-model (VM) \cite{vicsek_95_97,nagy_07} with $N$ point particles at number density $\rho$, which move at constant 
speed $v_0$.
The particles with
positions ${\bf x}_i(t)$ and velocities ${\bf v}_i(t)$
undergo discrete-time dynamics
with time step $\tau$. The evolution consists of two steps: streaming and (microscopic) collision.
In the streaming step
all positions are updated according to
\begin{equation}
\label{STREAM}
{\bf x}_i(t+\tau)={\bf x}_i(t)+\tau {\bf v}_i(t)\,.
\end{equation}
Because the particle speeds stay the same at all times, the velocities are parametrized by the ``flying'' angles, $\theta_i$,
${\bf v}_i=v_0(\cos{\theta_i},\sin{\theta_i})$.
In the collision step,
the directions $\theta_i$ 
are modified.
Particles align with their neighbours within a fixed distance $R$ plus some external noise:
a circle of radius $R$ is drawn around the focal particle $i$, and the average direction $\Phi_i$ of motion
of the
particles (including particle $i$)
within the circle is determined
according to
\begin{equation}
\label{COLLIS}
\Phi_i={\rm arctan}[\sum_{\{j\}} {\rm sin}(\theta_j)/\sum_{\{j\}} {\rm cos}(\theta_j)]\,,
\end{equation}
Eq. (\ref{COLLIS}) means that the vector sum of all particle velocities in every circle is computed and the direction of this summed vector
is taken as average angle $\Phi_i$.
Once all average directions $\Phi_i$ are known, the new directions
follow as
\begin{equation}
\label{VM_RULE}
\theta_i(t+\tau)=\Phi_i+\xi_i
\end{equation}
 where $\xi_i$ is a random number with zero mean and probability
distribution $w_n(\xi)$. The distributions are assumed to be even, $w_n(\xi)=w_n(-\xi)$ and normalized on the interval
$[-\pi,\pi]$ by $\int_{-\pi}^{\pi }w_n\,d\xi=1$.
In the original VM, $w_n(\xi)$ is a simple rectangular distribution, where 
$\xi$ is uniformly distributed in
the interval $[-\eta/2,\eta/2]$.
Here, 
I use a more general definition and assume that the shape of $w$ can depend on the number of particles enountered
in a collision circle. For example, 
self-interaction or simple diffusion where particle $i$ finds itself alone in a circle can 
be described by a different probability distribution
$w_1$ than binary collisions with distribution $w_2$.

The so-called standard Vicsek-model uses a forward-upating rule, as discussed in Ref. \cite{baglietto_08_09}.
The already updated positions ${\bf x}_i(t+\tau)$  
are used to determine the average directions $\Phi_i$.
In the so-called original VM, an Euler-like backward-updating rule is implemented. Here, the old locations ${\bf x}_i(t)$ are used 
to calculate the average directions $\Phi_i$.

\subsection{Continuous versus discrete time evolution}
\label{subsec:contin}

In gases with %
Hamiltonian dynamics, two important length scales immediately come to mind,
the effective range of interaction, $R$, and the average distance between molecules, $l_D=1/\sqrt{\rho}$
(all expressions in this paper are given for two dimensions).
Both lengths enter the well-known expression for the average distance particles travel between subsequent
binary
collisions,
\begin{equation}
\label{LCOLL}
l_{coll}={1\over \sqrt{2}\, \sigma\, \rho}={1\over 2\sqrt{2}} {l_D^2\over R}\,,
\end{equation}
with the 2D cross section $\sigma=2R$.
To solve the equations of motions numerically, a small time-step $\tau$ is introduced.
This creates a new length scale, $\lambda=v\,\tau$ which is irrelevant if it is much smaller than all other physical length scales,
$\lambda\ll {\rm min}(R,l_D,l_{coll})$. If this condition is met in the Vicsek-model, I will call this the {\em continuous-time VM}.
However, if the new scale $\lambda$ becomes larger or of the order of one of the previous lengths, $\lambda$ becomes relevant
and I label the model {\em discrete-time VM}.
It is easy to check that the original numerical work by Vicsek \cite{vicsek_95_97} was done in the continuous-time regime \cite{foot4}.
Why do we care about about the other regime?
One reason is that
kinetic and hydrodynamic theories are traditionally based on the smallness of some parameter.
For example, Boltzmann and later Bogolyubov \cite{bogo_46} exploited the smallness of the ratio $(R/l_D)^3$ at small density;
Landau and Vlasov \cite{vlasov_38} used the smallness of the interaction energy compared to the kinetic energy of molecules for their theories.
In our case, the new length scale $\lambda$ allows the definition of a different expansion parameter, $\vre=R/\lambda$.
The mean-field kinetic theory presented in the next chapter exploits the smallness of $\vre$, and, in fact, 
is actually the zeroth order contribution in a formal expansion in powers of $\vre$.
The advantage of introducing $\vre$ is that it allows to control the Molecular Chaos approximation.

For the VM defined by Eqs. (\ref{STREAM},\ref{VM_RULE}) the length $\lambda$ takes on the role of the {\em mean free path} (mfp) which is defined 
by the distance a particle travels between collision steps.
This is
because at every time increment $\tau$, particle directions will change, even if a 
particle has no collision partner and just undergoes self-interaction.

\section{Kinetic theory}
\label{sec:Kinetic}

In the VM, a given particle $i$ is specified by three numbers, its location $x_i$, $y_i$, and the flying angle $\theta_i$.
Hence, the microstate of a system of $N$ such particles is completely specified by $3N$ numbers
and corresponds to
a point in $3N$-dimensional phase space.
The time-evolution of the Vicsek model in this phase space is completely
Markovian because information about microstates from earlier times is irrelevant for further evolution.
This allows us to write down the Chapman-Kolmogorov equation for a Markov chain,
\begin{equation}
\label{MASTER1}
P({\bf B},t+\tau)=\int P({\bf A},t)\;W_{AB}\; d{\bf A}\,.
\end{equation}
where $P$ is the $N$-particle probability density \cite{foot1}.
Eq. (\ref{MASTER1}) 
describes the transition from a microscopic state ${\bf A}$
to the state ${\bf B}$ during one time step with transition probability $W_{AB}$.
The microscopic state of the system
at time $t+\tau$ is given by
the 3N-dimensional
vector,
$B\equiv(\theta^{(N)}, {\bf X}^{(N)})$, where
$\theta^{(N)}\equiv(\theta_1,\theta_2,\ldots, \theta_N)$ contains the flying directions of all $N$ particles, and
${\bf X}^{(N)}\equiv({\bf x}_1,{\bf x}_2,\ldots, {\bf x}_N)$ describes all particle positions.
The initial microscopic state at time $t$ is denoted as
${\bf A}\equiv(\tilde{\theta}^{(N)}, {\bf \tilde{X}}^{(N)})$.
The integral over the initial state translates to
$\int\,d{\bf A}\equiv \prod_{i=1}^N\int_{-\pi}^{\pi}\,d\tilde{\theta}_i\int\,d{\bf \tilde{x}}_i$
and ensures that all possibilities to create the state ${\bf B}$ are included. 
Pre-collisional angles and positions are given by $\tilde{\theta}_j$ and ${\bf \tilde{x}}_i$, respectively.
The transition probability $W_{AB}$ encodes the microscopic collision rules,
\begin{equation}
W_{AB}=
\prod_{i=1}^N
\delta({\bf \tilde{x}}_i-{\bf x}_i+\tau {\bf v}_i) 
\,\int_{-\pi}^{\pi}\, w_n(\xi_i)\;\,
\hat{\delta}(\theta_i-\xi_i-\Phi_i) 
\;\,d\xi_i\,, 
\label{TRANS1}
\end{equation}
and consists of two parts: the first $\delta$-function describes the streaming step
which changes particle positions. The second part contains
the periodically continued delta function,
$\hat{\delta}(x)=\sum_{m=-\infty}^{\infty}\delta(x+2\pi m)$,
which
accounts for the modification of angles in the collision step.
The particle velocities
${\bf V}^{(N)}\equiv({\bf v}_1,{\bf v}_2,..., {\bf v}_N)$, are given in terms
of angle variables $\theta_i$, 
\begin{equation}
{\bf v}_i=v_0\;{\bf e}_i(\theta)=v_0\,(\cos{\theta_i}, \sin{\theta_i})\,.
\end{equation}
with unit velocity vectors ${\bf e}_i$.
For the standard VM, a flat noise distribution with noise strength 
$\eta$ is used that does not depend on the actual particle number $n$ in the interaction
circle,
\begin{equation}
\label{ETADEF}
 w_n(\xi) = \begin{dcases*}
        {1\over \eta}     & for $-{\eta\over 2}\le \xi \le {\eta\over 2}$\\
        0                 & elsewhere.
        \end{dcases*}
\end{equation}
Note that Eq. (\ref{TRANS1}) corresponds to the forward-update rule (standard VM) 
\cite{baglietto_08_09} as used in the agent-based simulations of Ref. \cite{chate_04_08}.
Results for backward-updating will be given elsewhere.

Equation (\ref{MASTER1}) 
can be interpreted as the discrete time analogue of the 
Liouville equation of statistical mechanics.
It is
exact but
intractable without simplification.
The easiest way to proceed is to make Boltzmann's molecular chaos approximation
by assuming that the particles are uncorrelated
just prior to a every microscopic interaction,
which amounts to a factorization of the N-particle probability into a product of one-particle probabilities,
$P(\theta^{(N)}, {\bf X}^{(N)})
=\prod_{i=1}^N P_1(\theta_i, {\bf x}_i)$ on the right hand side of Eq. (\ref{MASTER1}).
This approximation 
is useful at sufficiently large noise strength and
when the mean free path (mfp)
is large
compared to the radius of interaction $R$.
As discussed in Chapter \ref{subsec:contin}, the mfp is given by the distance a particle travels between collision steps,
$\lambda=\tau\,v_0$. 
This expression differs from the usual density-dependent formula in regular gases, Eq. (\ref{LCOLL}),
because of the special discrete time dynamics of the VM and 
the fact that particles in the VM do not interact during streaming.
A large mfp and sufficiently large noise ensure that particles are well mixed and that the probability of
subsequent re-collisions of the same particles is small, which supresses memory and correlation effects.
These correlations only vanish completely in extreme limits, for example, 
when the noise $\eta$, defined in Eq. (\ref{ETADEF}), is exactly equal to $2\pi$ and particles just diffuse but do not interact at 
all,
or when the mean free path and the system size are infinite while $\eta$ is nonzero.
However, even under more realistic conditions, the molecular chaos assumption can lead to very accurate results, 
see for example,
Refs. \cite{ihle_03_05,pool_05,ihle_13}. 
Because molecular chaos neglects
pre-collisional correlations and leads to an effective one-particle picture,
the final outcome will be a mean-field theory.
So far, to my knowledge, all treatments of active particles
by kinetic theory, for example \cite{bussemaker_97,bertin_09,mishra_10,mishra_12,bertin_13,solon_13},
or Fokker-Planck equations \cite{romanczuk_12,grossmann_13} make such a mean-field assumption, either explicitly or
implicitly.

To derive this mean-field theory, we multiply eq. (\ref{MASTER1}) by
the phase space density $\sum_i\delta({\bf v}-{\bf v}_i)\delta({\bf x}-{\bf x}_i)$, \cite{foot2}.
A subsequent integration
over all particle positions $x_i$ and angles $\theta_i$ leads,
in the large $N$-limit,
to a kinetic equation for
the one-particle distribution function,
$f(\theta,{\bf x},t)=NP_1(\theta,{\bf x},t)$ 
\begin{equation}
\label{VLASOV}
f(\theta, {\bf x}+\tau {\bf v},t+\tau)=
\int_0^{2\pi}f(\phi, {\bf x},t)\; G({\bf x},\theta,\phi,t)
\;d\phi
\end{equation}
with the nonlocal (in velocity space) mean-field potential $G$ 
that acts like an external potential. 
Particles are assumed to move in an uncorrelated fashion and the effect of their mutual interactions being such that any one
particle experiences the average potential field $G$ that depends nonlinearly and nonlocally  on $f$ itself.
This picture is thus similar to the Vlasov kinetic equation, the Hartree and the Debye-H{\"u}ckel theory.
The potential $G$ 
is given by
\begin{eqnarray}
\nonumber
& &G({\bf x},\theta,\tilde{\theta}_1,t)=
\bigg\langle \bigg\langle
\sum_{n=1}^N
\int_{-\pi}^{\pi}
w_n(\xi)\,
d\xi\,
{{\rm e}^{-M_R}\over (n-1)!}
\\
\label{ENSKOG1}
& &\times
\,\hat{\delta}[\theta-\xi-\Phi_1(\tilde{\theta}_1,\ldots\tilde{\theta}_n)]
\,\prod_{i=2}^n f( \tilde{\theta}_i, {\bf x}_i,t)
\bigg\rangle_{\tilde{\theta}} \bigg\rangle_x
\end{eqnarray}
where
$M_R({\bf x},t)=\int_R\rho({\bf y},t)\,d{\bf y}$
is the average number of particles in a circle of radius $R$ centered around ${\bf x}$ and
can be position dependent.
The subscript ``R'' at the integral denotes integration over this circle.
The local particle density $\rho$ is given as a moment of the distribution function,
$\rho({\bf x},t)=\int_0^{2\pi} f(\theta,{\bf x},t)\,d\theta$;
$\langle ... \rangle_x=\int_R... \,d{\bf x}_2\,d{\bf x}_3...d{\bf x}_n$
denotes the integration over all positions, $n-1$ particles can assume within the interaction circle;
$\langle ... \rangle_{\tilde{\theta}}=\int_{-\pi}^{\pi} ... d \tilde{\theta}_2 d \tilde{\theta}_3 ...
d \tilde{\theta}_n $
is the average over the pre-collisional angles
of all particles in the interaction circle except particle $1$.
Here, particle $1$ is assumed to be the focal particle. 
It is fixed at position ${\bf x}$ and particles $2,3\ldots n$ are supposed to be its neighbors.
Of course, this is not the only possibility but since the particles are identical, all particle permutations 
give the same contribution and are already taken into account by 
the combinatorial factor $1/(n-1)!$. 
It is interesting to note that this combinatorial factor together with the exponential ${\rm e}^{-M_R}$
describes a Poisson distribution of the particle locations.
The Poissonian character of these fluctuations was not put in by hand -- it rather is a consequence
of the Molecular Chaos approximation and the definition of $f(\theta,{\bf x})$ as the ensemble average
of the microscopic phase space density.
That means, the density fluctuations of an ideal gas are already implicitly contained in Eq. (\ref{VLASOV}).
Therefore, one has to be very careful with inserting additional noise terms to construct fluctuating 
kinetic equations,
something
which is quite popular for other systems, \cite{bixon_69,gross_10,kaehler_13}.
Note, that the exponential prefactor ${\rm exp}(-M)$ is only abtained in the thermodynamic limit, $N\rightarrow \infty$.
In realistic active particle systems, $N$ is not that large. However, even for finite $N$ a version of Eq. (\ref{ENSKOG1}) can be derived
where the Poisson distribution is replaced by a binomial factor \cite{chou_14} in order to better describe agent-based simulations
with small $N$.

\section{Phase diagram and hydrodynamic equations}
\label{sec:Phase_Hydro}

\subsection{Calculating the phase diagram}
\label{subsec:Phase}

For stationary and spatially homogeneous solutions,
the mean-field kinetic equation (\ref{VLASOV}) 
turns into a nonlinear Fredholm integral equation of the second kind,
\begin{eqnarray}
\nonumber   
& &f(\theta)=
\sum_{n=1}^N\,
\int_{-\pi}^{\pi}w_n(\xi)\,
d\xi\,
{A^{n-1} {\rm e}^{-M}\over (n-1)!}
\\
\label{FRED}
&\times & \bigg ( \prod_{i=1}^n \int_{-\pi}^{\pi}f(\tilde{\theta}_i)\,d\tilde{\theta}_i\bigg )
\,\hat{\delta}[\theta-\xi-\Phi_1(\tilde{\theta}_1,\ldots\tilde{\theta}_n)]
\end{eqnarray}
where $A=\pi R^2$ is the area of the collision circle and the average particle number in this circle, $M=A\rho_0$, is proportional to
the particle number density $\rho_0$.
It is easy to see that the constant distribution $f_0=\rho_0/(2\pi)$ that describes a disordered state
is a solution for all possible noise distributions.
Ordered solutions $f(\theta)\neq constant$ with nonvanishing polar order can be determined numerically 
and bifurcate continuously from the disordered solution.
For the noise distribution defined in Eq. (\ref{ETADEF}) one finds that the critical noise below which
the ordered state exists, 
follows from the
condition $\Lambda=1$,
where $\Lambda$ is defined as
\begin{eqnarray}
\nonumber
\Lambda&=&{4 \over \eta}{\rm sin}\left({\eta\over 2}\right)
{\rm e}^{-M} \sum_{n=1}^N {n^2 M^{n-1}\over n!}\,I(n) \\
\label{LAMBDA_DEF}
I(n)&=&{1\over (2\pi)^n}
\int_0^{2\pi}d\theta_1\ldots
\int_0^{2\pi}d\theta_n
\,\cos{\theta_1}
\,\cos{\Phi_1(\theta_1,\ldots\theta_n)}\,
\end{eqnarray}
Here, 
$\Phi_1$ is the average angle defined in Eq. (\ref{COLLIS}).
The integral $I(n)$ and similar integrals were evaluated analytically for $n\le 3$ and numerically for $n\le 10$.
In addition, asymptotic expressions for $n\rightarrow \infty$ are known \cite{ihle_11,chou_12}.
For values $n>10$ an interpolation between the known integrals at low $n$ and the asymptotic results was used.

Analyzing the condition $\Lambda=1$ in the low density limit $M\ll 1$, using $I(1)=1/2$ and $I(2)=1/\pi$, leads to
an explicit expression for $\eta_C$,
\begin{equation}
\label{ETAC_REGULAR_VM}
\eta_C=\sqrt{48 M\left( {2\over \pi}-{1\over 2}\right)}
+O(M)\,,
\end{equation}
In the opposite limit of infinite density, $\eta_C$ goes to $2\pi$. For the behavior at intermediate densities, see Ref. 
\cite{ihle_11}.
Close to the bifurcation, that is at $\eta_c-\eta\ll 1$, an analytical solution of Eq. (\ref{FRED}) 
in terms of angular Fourier modes
can be obtained and, as expected, one finds that the order parameter of the state of collective motion
follows the mean-field scaling $\Omega\sim (\eta_c-\eta)^{1/2}$.
It turns out \cite{bertin_06,bertin_09,ihle_11}, 
that the homogenoues ordered state just near $\eta_C$ is linearly unstable to long wavelength fluctuations,
at least for large mean free path, $v_0\tau/R\gg 1$.
These perturbations turn into steep  soliton-like waves \cite{chate_04_08}.
Direct simulations of Eq. (\ref{VLASOV})
demonstrate that these waves
show hysteresis and turn the flocking transition
into a discontinuous phase transition \cite{ihle_13}.
Recently, Thueroff {\em et al.} \cite{thueroff_13} 
observed similar wave behavior by directly simulating the Boltzmann equation 
proposed by Bertin {\rm et al.} \cite{bertin_06,bertin_09}.
Since the PSA approach is only valid at large mfp and a Boltzmann approach is in practice never valid 
near the transition
as shown in Sect.~\ref{sec:Escape}, it is not clear yet whether the same soliton-scenario
applies also at the highly correlated regime of small mfp, $v_0 \tau/R\ll 1$.
In this low velocity regime, see for example Fig. 1 of Ref. \cite{nagy_07}, isolated flocks were observed
rather than the straight, boundary-spanning density waves of the high velocity regime, shown in Figs. 4 and 5 of Ref. \cite{nagy_07}.

\subsection{Deriving hydrodynamic equations}
\label{subsec:Hydro}

The Chapman-Enskog expansion (CE) from 1916 is one of the standard techniques to extract macroscopic behavior
from kinetic equations \cite{chapman_70}.
It can be seen as an elaborate expansion in small gradients of hydrodynamic 
fields.
Its key assumption is that after a few collisions that involve rapid changes of the distribution function $f$, 
the system reaches a ``hydrodynamic state''
where local equilibrium is achieved and
where $f$ can be expressed as a functional of the slow hydrodynamic variables \cite{foot3}.
This means, $f$ is expected to depend on space and time only indirectly through those hydrodynamic fields. 
The hydrodynamic variables are just the lowest velocity moments of $f$, for example, density $\rho$ and momentum density
${\bf w}$
are given by
\begin{eqnarray}
\nonumber
\rho({\bf x},t)&=&\int_{-\pi}^{\pi}f({\bf x},\theta,t)\;d\theta \\
\label{MOMENT_DEF}
{\bf w}({\bf x},t)=\rho{\bf u}&=&\int_{-\pi}^{\pi}{\bf v}(\theta)\;f({\bf x},\theta,t)\;d\theta 
\end{eqnarray}
The Chapman-Enskog assumption can be rephrased as the claim that knowledge
of just the first few moments of $f$ is sufficient to describe the system on large length and time scales. 
Since $f$ is uniquely defined by all its moments, this assumption would be justified if either all higher moments are 
negligibly small or that they are ``enslaved'' to the lower moments, meaning that they could be expressed as functionals
of the lower moments. 

Applying CE to the Vicsek model is tricky because the only true and nontrivial hydrodynamic field is the density $\rho({\bf x},t)$.
This is because momentum is typically not conserved by the collision rule, Eq. ({\ref{VM_RULE}), and
energy is trivially conserved since the particle speed is constant. However, at the order-disorder threshold the interplay
between angular noise and alignment leads to momentum conservation in an averaged sense.
Mathematically, this can be seen in the evolution equation for momentum density, Eq. (\ref{FINAL_EQ}), 
a Navier-Stokes-like equation,
which has a gain/loss term that vanishs at the threshold. Therefore, in the VM, I will treat momentum density as a pseudo-hydrodynamic mode.
Since we are mostly interested in the behavior near the threshold, we should have a separate equation for this variable;
that is exactly what Toner and Tu \cite{toner_98} postulated -- one equation for the density and one for the momentum density.

The Chapman-Enskog expansion takes
the local stationary state as a reference state and expands
around it in powers of the hydrodynamic gradients.
To systematically account for these gradients 
a dimensionless ordering parameter $\epsilon$ is introduced, which is set to unity at the end
of the calculation. The physical meaning of this parameter is that it 
assumed to be proportional to the Knudsen number, e.g. 
the ratio of the mean free path to the length scale over which hydrodynamic fields change considerably.
The CE procedure  
starts with 
a Taylor expansion of the l.h.s of Eq. (\ref{VLASOV})
around $(\theta, {\bf x}, t)$.
The spatial gradients that occur are scaled as
$\partial_{\alpha}\rightarrow \epsilon\partial_{\alpha}$, 
and
multiple time scales $t_i$ are introduced in the temporal gradients.
For the VM, the following scaling that respects the physics of the microscopic collisions was chosen,
\begin{equation}
\partial_t=\partial_{t_0}+\epsilon\partial_{t_1}+\epsilon^2\partial_{t_2} \ldots\, .
\end{equation}
This sequence differs from the usual set of equations for models with momentum conservation \cite{mcnamara_93,ihle_00}
because of the fast time scale $t_0$ which is not multiplied by a power of $\epsilon$ and
contributes time derivatives of all orders. 
However, expansions that
contain all powers of $\partial_{t_0}$
can be conviniently summed up by the time evolution operator
\begin{equation}
\label{TIME_EVOLV0}
T={\rm exp}\left(\tau \partial_{t_0}\right)
\end{equation}
which shifts the time-argument of a function by the discrete time step $\tau$,
$T\circ f(t)=f(t+\tau)+O(\epsilon)$.

The next step in the CE is to expand the distribution function $f$ and the collision integral $C$, e.g. the right hand side of Eq. (\ref{VLASOV}), in powers 
of $\epsilon$, 
\begin{eqnarray}
\nonumber
f&=&f_0+\epsilon f_1 +\epsilon^2 f_2+\ldots \\ 
C&=&C_0+\epsilon C_1+\epsilon^2 C_2+\ldots\,. 
\end{eqnarray}
Inserting this into 
Eqs.\ (\ref{VLASOV}, \ref{ENSKOG1}), and collecting terms of the same order
in $\epsilon$ yields a hierarchy of evolution equations for the $f_i$.
Due to the absence of momentum conservation and Galilean invariance this
set 
of equations is very different from the usual one. 
It is not {\em a priori} evident whether the scaling ansatz 
for the time derivatives is
correct.
However, it turns out that this choice 
avoids any inconsistencies 
if additionally 
the expansion of the distribution function $f$ is
identified as an angular Fourier series
with 
$f_0({\bf x},t)=\rho({\bf x},t)/ 2\pi$ and, for $n>0$,
$f_n({\bf x},\theta,t)=\left[a_n({\bf x},t)\cos{(n\theta)}+
b_n({\bf x},t)\sin{(n\theta)}\right]/\pi v_0^n
$.

The goal is to find a hydrodynamic description of the first two moments of $f$, namely the particle density
and the macroscopic momentum density.
Inserting the Fourier representation of $f$ into the definition of these moments, Eqs. (\ref{MOMENT_DEF}), shows that 
the coefficients for the first order contribution $f_1$ are given by the momentum density, $a_1=w_x$ and $b_1=w_y$. 
Multiplying the hierarchy of evolution equations by powers of the microscopic velocity
vector $\vec{v}=(v_x, v_y)=v_0({\rm cos}\theta, {\rm sin}\theta)$ and integrating over $\theta$ gives a set of equations
for the time development of the density and the moments $a_i$ and $b_i$.
These equations still depend on higher order moments. %
To significantly simplify the closure of this hierarchy of moment equations,
the analysis is restricted 
to the vicinity of the threshold where $\Lambda$, defined in eq. (\ref{LAMBDA_DEF}), is close to one. 
Specifically, I assume the scaling, $1-\Lambda=O(\epsilon^2)$.
This allows me to express the time evolution of the moments of the higher order distribution functions
$f_2$ and $f_3$ in terms of gradients of the hydrodynamic fields. This means these functions depend on time only implicitly
through their functional dependence on $f_0$ and $f_1$.
Thus, at order $O(\epsilon^3)$ and near the flocking threshold, I found that the moments $f_2$ and $f_3$ are enslaved
to $f_0$ and $f_1$, whereas even higher functions such as $f_4$ can be neglected at this order.
This results in a consistent closure of the hierarchy equations and leads to two hydrodynamic equations with a larger number of terms than postulated \cite{toner_98} or derived by other authors \cite{bertin_09}.
The question that come to mind is, what would happen if the system is further away from the threshold where $1-\Lambda$ is not small? It is possible that two hydrodynamic equations will not be sufficient anymore. Equations for higher 
order (non-hydrodynamic) moments might be needed or a description in more convinient variables might be more useful.

All equations are rescaled by expressing time 
in units of $\tau$ and distances in units of the mfp, $\tau v_0$,
which also makes $\rho$ and $\vec{w}$ dimensionless.
After tedious calculations one obtains
the continuity equation $\partial_t\rho+\partial_{\alpha}
w_{\alpha}=0$, and an equation for the 
momentum density,
\begin{equation}
\label{FINAL_EQ}
\partial_{t}\vec{w}+\nabla\cdot\HT=-b\nabla\rho+(\Lambda-1)\vec{w}+
\QTa\cdot \vec{w}+\QTb\cdot\nabla\rho
\end{equation}
with $b=(3-\Lambda)/4$.
The momentum flux tensor $\HT$ and the tensors $\QTa$, $\QTb$,
\begin{equation}
\label{TENSOR1}
\HT=\sum_{i=1}^5h_i\,\omi\;\;\;\;\;\;\;
\QTa=\sum_{i=1}^5q_i\,\omi \;\;\;\;\;\; 
\QTb=\sum_{i=1}^5k_i\,\omi
\end{equation}
are given in terms of five symmetric traceless tensors $\omi$, 
\begin{eqnarray}
\nonumber
\Omega_{1,\alpha\beta}&=&\partial_{\alpha}w_{\beta}+\partial_{\beta}w_{\alpha}
-\delta_{\alpha\beta}\partial_{\gamma}w_{\gamma} \\
\nonumber
\Omega_{2,\alpha\beta}&=&2\partial_{\alpha}\partial_{\beta}\rho
-\delta_{\alpha\beta}\partial^2_{\gamma}\rho \\
\nonumber
\Omega_{3,\alpha\beta}&=&2w_{\alpha}w_{\beta}
-\delta_{\alpha\beta}w^2 \\ 
\nonumber
\Omega_{4,\alpha\beta}&=&w_{\alpha}\partial_{\beta}\rho+w_{\beta}\partial_{\alpha}\rho
-\delta_{\alpha\beta}w_{\gamma}\partial_{\gamma}\rho \\
\label{OMEGA_DEF}
\Omega_{5,\alpha\beta}&=&2(\partial_{\alpha}\rho)(\partial_{\beta}\rho)
-\delta_{\alpha\beta}(\partial_{\gamma})^2\,,
\end{eqnarray}
which are all of order O($\ve^2$).
The tensor $\Omega_1$ is the viscous stress tensor of a two-dimensional fluid.
The transport coefficients $h_i$, $k_i$ and $q_i$ 
were explicitly obtained in the limit of large mfp, $\tau\,v_0\gg R$, mainly
for simplicity but also because the PSA approach is not expected to be reliable at low mfp.
The detailed expressions are given in Ref. \cite{ihle_11}. They are valid at arbitrary density and depend
on $v_0\tau/R$, $\eta$ and $M$. The concerns expressed in Ref. \cite{peshkov_thesis} on the complexity of these expressions
are only partially justified because the occuring sums and angular integrals can be quite accurately evaluated using a 
Mathematica$\textsuperscript{\textregistered}$ script and
the interpolation 
technique mentioned in the discussion of Eq. (\ref{LAMBDA_DEF}). %

Expressing the Navier-Stokes-like equation, Eq. (\ref{FINAL_EQ}), in terms of tensors and vectors makes it easier to
see that all terms are rotationally invariant. This is because once one has verified that 
all the $\omi$ transform like tensors, the products of these quantities with vector like $\nabla \rho$ or ${\bf w}$ also
transform like vectors and thus are rotationally invariant.
One also sees now explicitly that the loss-term that is linear in ${\bf w}$ has the prefactor $\Lambda-1$ and therefore
vanishs at
the threshold. That means at the threshold, macroscopic momentum 
is approximately conserved.

Comparing Eq. (\ref{FINAL_EQ}) with the equations postulated in Ref. \cite{toner_98} and amended 
by Toner \cite{toner_12} there seems to be additional terms that are not included in the Toner-Tu theory.
For example, analyzing the x-component of Eq. (\ref{FINAL_EQ}) one finds a contribution
$w_x(\partial_x^2\rho-\partial_y^2\rho)+2w_y\partial_x\partial_y\rho)$
in tensor $Q_1$ which originates from the tensor $\Omega_2$. 
This term is relevant for the linear stability of the ordered state.
No term in Eq. (1) of Ref. \cite{toner_12}
is able to reproduce this expression, even if one assumes that the coefficients in that equation 
depend on density.
Because these novel terms have not been systematically analyzed yet, it is not clear 
whether they change anything in the main conclusions of the
Toner-Tu theory.

A recent derivation of hydrodynamic equations by Gro{\ss}mann {\rm et al.} \cite{grossmann_13} for active Brownian particles leads to similar
terms.
By assigning a power of $\ve$ to every gradient and to every occurence of the momentum density, ${\bf w}\sim \ve$, one can write down
all possible products of $\rho$, ${\bf w}$ and its gradients that are at most of order $O(\ve^3)$.
It appears that Eq. (\ref{FINAL_EQ}) contains all possible, rotationally invariant terms of this type (assuming that the coefficients
of the terms are density-dependent).
Therefore, I believe that models with the same symmetries as the VM such as the metric-free model of 
Refs. \cite{chate_10,ngo_12,chou_12}
will lead to hydrodynamic equation with the same terms just with different coefficients.

\subsection{Validity of the hydrodynamic equations}
\label{subsec:Verify}

So far, a direct term-by-term verification of the hydrodynamic equations, Eq. (\ref{FINAL_EQ}), 
is lacking. 
However, a number of indirect consistency tests were successfully completed.
One of them was the numerical solution of the kinetic equation, Eq. (\ref{VLASOV}), in large systems \cite{ihle_13}.
This test, at least, probes the kinetic foundation from which the hydrodynamic description was derived
but, of course, cannot prove the validity of the hydrodynamic equations.
The shape and speed of the solitons observed in these runs agreed within a few percent with agent-based simulations of the VM, at $\tau v_0/R\gg 1$.
To check the hydrodynamic equations, we have also performed a linear stability analysis of both the hydrodynamic equation, Eq. (\ref{FINAL_EQ}), in Ref. \cite{ihle_11}, and the kinetic equation, see Ref. \cite{chou_12}.
Since the hydrodynamic equations are only supposed to be valid close to the threshold and the analysis of the kinetic equation
was done in the binary collision approximation, we chose $\eta_C-\eta\ll 1$ and $M\ll 1$ where both approaches should be valid.
In this limit, the dispersion relations for a small longitudinal perturbation of the ordered state, agreed quantitatively 
with each other
\cite{chou_14}.
Finally, the numerical solution of the hydrodynamic equation by a finite-difference scheme showed a 
linear instability of the ordered phase once the system size exeeds a certain length. A length of similar size above which spatial 
inhomogeneities occured 
was found in simulations of the kinetic equation. 

A well-known issue is the possible divergence of higher order Chapman-Enskog and similar gradient expansions \cite{karlin_02}.
The most famous example is that in regular gases an expansion to second order gives the  
stable Navier-Stokes equation but going to the next order leads to the Burnett equation that turns out to be unstable \cite{bobylev_82}. 
For the VM, a similar problem seems to occur. 
The numerical solution of Eq. (\ref{FINAL_EQ}) correctly shows a linear long wavelength instability once the system size exceed a 
critical length at slightly smaller noise than the threshold noise \cite{ihle_11}. 
However, in contrast to agent-based simulations and direct simulations of the kinetic equation, these perturbations never settle but keep growing, e.g. are nonlinearly unstable.
This means, the hydrodynamic equations, Eqs. (\ref{FINAL_EQ})-(\ref{OMEGA_DEF}), whose coefficients were all derived from the microscopic rules, were unable to reproduce stable, inhomogeneous solutions.
Some researchers handle this problem by inserting one or more higher order nonlinearities phenomenologically to control 
this behavior \cite{mishra_10},
others perform a tedious summation to all orders \cite{karlin_02,slemrod_12} or use 
non-perturbative techniques 
similar to the 
Schwinger-Dyson equation \cite{karlin_14}.
However, the latter techniques, while very promising, might only be feasible 
for simpler systems than the VM. 
Such summations of higher order gradients and nonlinearities can well lead to nonlocal hydrodynamics \cite{slemrod_12}.
A pragmatic solution was presented in Ref. \cite{ihle_13} where, 
instead of dealing with complicated gradient expansions, the non-local kinetic equation (\ref{VLASOV})
was solved on the computer as is. This amounts to an implicit summation of gradient terms 
to all orders.

Another idea is to go to the next higher order, $O(\epsilon^4)$, in the CE, in the hope that the new nonlinearities behave ``nicer'' and regularize the instability 
but this would be 
very
tedious, the number of terms would become huge, and in the end, the extended equation might even be more unstable.
In Ref. \cite{peshkov_thesis} it was hypothesized that the nonlinear instability of Eq. (\ref{FINAL_EQ}) 
might be due to an incorrect
closure of the moment hierarchy. While this has not been completely ruled out yet for 
most kinetic theories of active matter,
I think that similar to the Burnett-equation one should rather {\em expect} a 
higher order gradient expansion of a kinetic equation to diverge at some point. 
I tend to believe that if such an expansion would be fully stable at all wavelengths and perturbation sizes, 
it must be 
a very lucky case. 

\section{Binary escape time and the failure of Boltzmann approaches}
\label{sec:Escape}
The Boltzmann equation is the most common kinetic equation for regular gases, and 
became also quite popular in active matter research \cite{bertin_06,peshkov_12,thueroff_13,bertin_13}.
The success of the Boltzmann approach for regular matter is based on 
its accuracy at
low densities.
The derivation of the Boltzmann equation involves a number of assumptions, the most important ones being
the Molecular Chaos assumption and the binary collision assumption, e.g. the neglect of collisions involving more than 
two particles.
It is an important question whether these crucial assumptions also hold in active matter systems.
Recently, the validity of the Molecular Chaos assumption (MC) for a realistic active colloidal system has 
been critically
assessed in Ref. \cite{hanke_13}. 
Indicators of molecular chaos were also investigated for the topological Vicsek-model \cite{chou_12}.
Here, I would like to focus on the binary collision assumption. %
\begin{figure}
\centering
\includegraphics[width=10.0cm,clip]{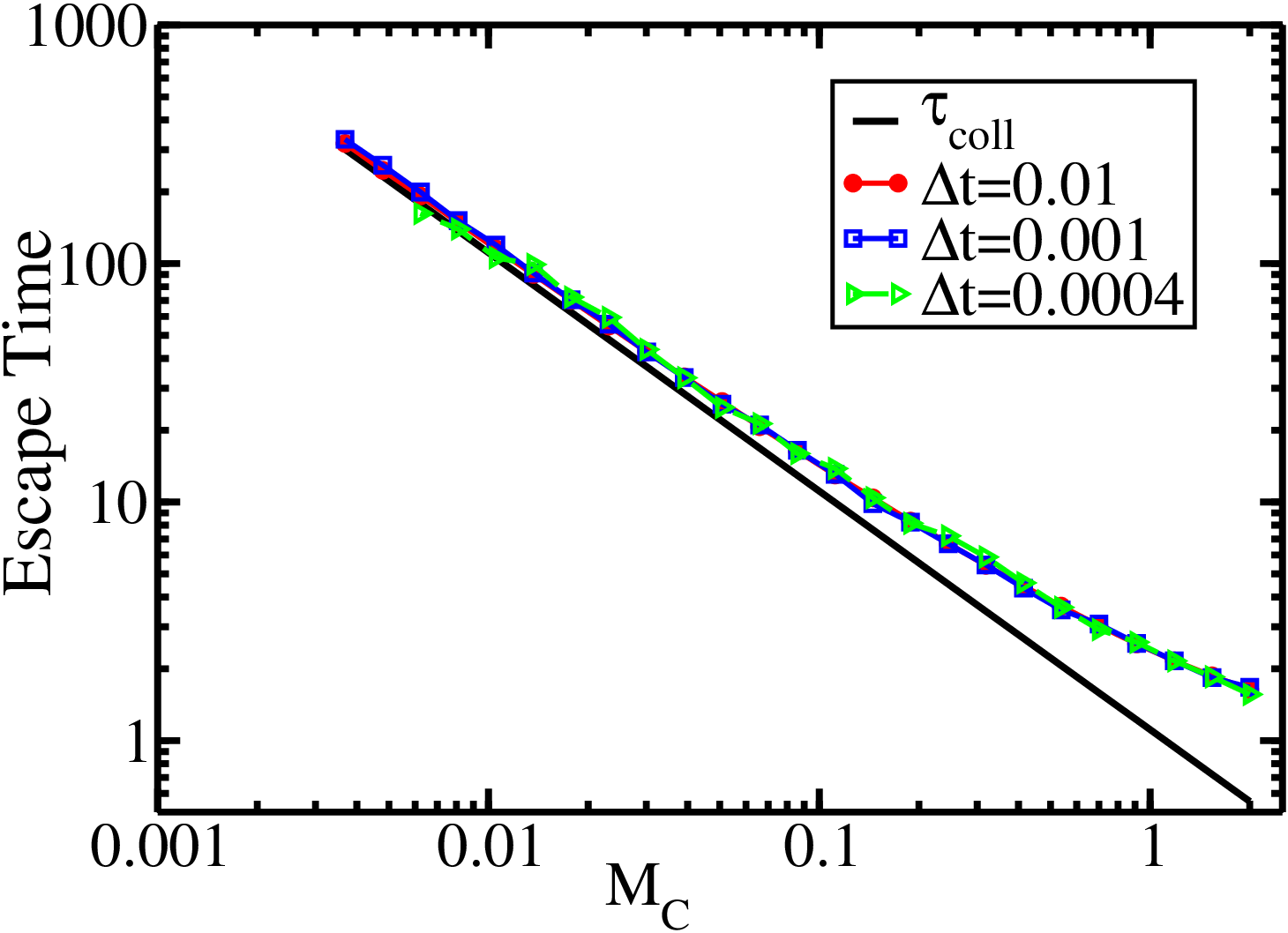}
\caption{The binary escape time $\tau_0$ measured in Monte-Carlo simulations as
a function of the threshold density $M_C=M(\eta_C)$ for different time steps $\Delta t$, averaged over $10^6$ runs.
The noise $\eta$ is set equal to
the mean-field critical noise given in Eq. (\ref{ETAC_REGULAR_VM}).
The solid black line is the time between collisions for a regular gas, $\tau_{coll}=l_{coll}/v_0$, see Eq. (\ref{LCOLL}).
Parameters: $R=1$, $v_0=1$.}
\label{fig:escape1}
\end{figure}

Lowering the density reduces 
the probability of non-binary collisions
in regular gases with short-ranged repulsion. 
This is because at low density 
the range of interaction $R$ is much smaller than the average distance between particles $l_D\sim 1/\sqrt{\rho}$.
According to Eq. (\ref{LCOLL}) this leads to $R\ll l_D \ll l_{coll}$ where $l_{coll}$ is the average distance a molecule 
travels until it collides with another one.
This disparity in length scales translates to the relevant time scales: The typical time a particle is engaged
in an interaction, $\tau_0\approx R/v$, where $v$ is the thermal speed, is much smaller than the time
between collisions, $\tau_{coll}$. This makes it very unlikely that a particle meets two or more
 other ones in its action radius within a short time interval of order $\tau_0$.
This scenario changes for attractive interactions, particles could capture each other and orbit around one
another. The alignment interactions of the VM, Eq. (\ref{VM_RULE}), can also have such a capturing effect at low
noise because particles move almost in parallel after a collision and have a tendency to stay together,
effectively prolonging the collision time $\tau_0$. If $\tau_0$ is large, the likelihood of a third particle
to join an ongoing binary encounter increases and three-particle interactions might become non-negligible.

As a first step to check the validity of the binary collision assumption, I measure
the collision time $\tau_0$ for the binary alignment interactions of the VM
by Monte Carlo simulations. 
These simulations involve only two particles which are initially placed at distance $r=R-2\lambda\,\xi$
where $\xi$ is a random number equally distributed in the interval $(0,1]$ and $\lambda=v_0\,\tau$.
This range of $r$ covers all possible positions two particles with $\lambda/R<1$ can have after they have entered
into each other's action circle for the first time.
Then, the initial flying directions, $\theta_1$ and $\theta_2$ are chosen randomly, 
and define the velocities of the particles.
To check the consistency of the intial conditions, the particles are traced back to their previous 
positions ${\bf x}_{i,old}={\bf x}_i-\tau {\bf v}_i$. If $d_{12}=|{\bf x}_{2,old}-{\bf x}_{1,old}|$
is smaller than the interaction radius $R$, the particles were not at the very beginning of a binary encounter.
In this case, the intial condition is discarded and a new set of positions and velocities is chosen
until the condition $d_{12}\ge R$ is met. 
Then, the two particles evolve according to
the collision and streaming rules of the VM, see Eqs.
(\ref{STREAM},\ref{VM_RULE}),
until their distance $d_{12}$ exceeds again the collision radius.
The time until this happens is the duration of a collision and will also be called {\em binary escape time}.
These measurement are repeated for different noises $\eta$ and different time steps,
and are ensemble-averaged over many random initial conditions.

Since kinetic theories for active particles are designed to describe the order-disorder transition and the state of
collective motion, the noise $\eta$ has to be close to the threshold noise $\eta_C$ for any interesting 
application.
Therefore, in order to estimate $\tau_0$ near the flocking threshold, 
I map the noise used in the simulations to the rescaled density
$M=\pi R^2 \rho$ using the mean-field expression, Eq. (\ref{ETAC_REGULAR_VM}), for
the threshold noise $\eta_C$. 
The average binary escape time $\tau_0$ is plotted as a function of $M_C=M(\eta_C)$ in Fig. \ref{fig:escape1}.
As seen in this figure, $\tau_0$ converges for time steps $\Delta t\le 0.01R/v_0$.
The main result is that, for small $M$, the escape time scales as $\tau_0\sim M^{-\nu}$ with an exponent $\nu\approx 1$, 
assuming the particular mapping between density and noise, $\eta=\eta_C\sim\sqrt{M}$.
For comparison, the time between collisions in a regular gas, $\tau_{coll}=l_{coll}/v_0\sim M^{-1}$ is also plotted.
While isolated particles in the VM do not go on straight lines like in regular gases but 
undergo a correlated random-walk due to self-interactions, I still expect
a similar scaling of $\tau_{coll}$ with $M$.
For small densities $M\le 0.1$, both times seem to be very close to each other and actually seem to follow nearly the same
scaling with density. One has to keep in mind that the observed $\tau_0$ likely 
overestimates the duration of a binary collision in the VM 
because the Monte Carlo procedure assumes 
that particle velocities are completely uncorrelated before
they enter each others action circle and only get strongly correlated while engaged in the collision.
This leaves out situations where particles that have just left each other recollide again while their directions 
did not have enough time to become very different from each other.
Nevertheless, my numerical results suggest that  
the ratio of $\tau_0$ and $\tau_{coll}$ remains of order one or at least
goes down very slowly with decreasing density.
This would mean, that contrary to regular gases the binary collision assumption 
does not become valid at small densities. 

\begin{figure}
\centering
\includegraphics[width=10.0cm,clip]{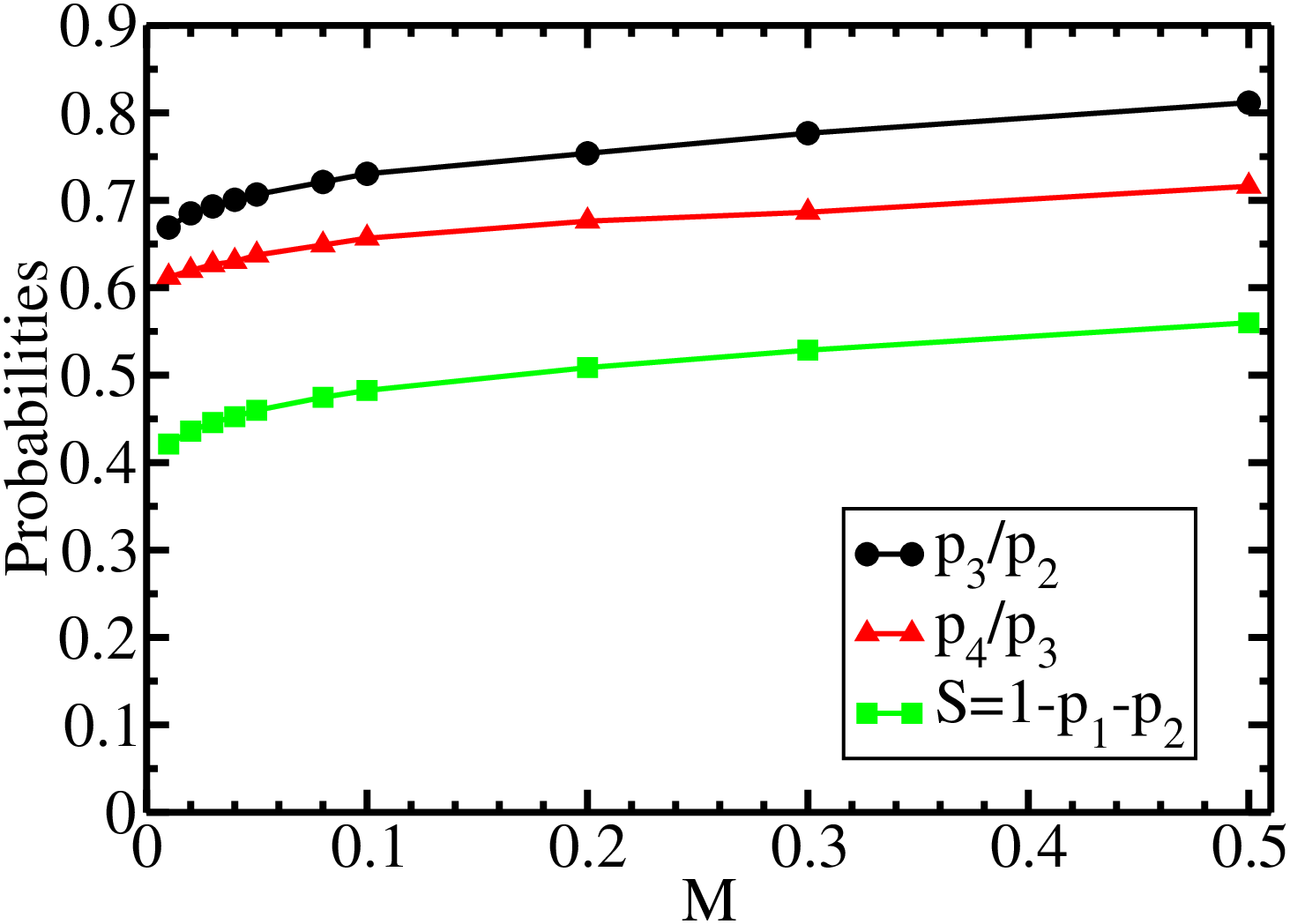}
\caption{Agent-based simulations of the probability ratios $p_3/p_2$, $p_4/p_3$, and the probability for non-binary interactions, $S=1-(p_1+p_2)$,
versus density $M$ at fixed time step $\tau=0.03$. The noise $\eta$ is set equal to the mean-field prediction $\eta_C(M)$ from Eq.
(\ref{ETAC_REGULAR_VM}).
Parameters: $R=v_0=1$, $N=6000$, up to $7.3\times 10^6$ time steps.
}
\label{fig:AGENT_vs_M}
\end{figure}
\begin{figure}
\centering
\includegraphics[width=10.0cm,clip]{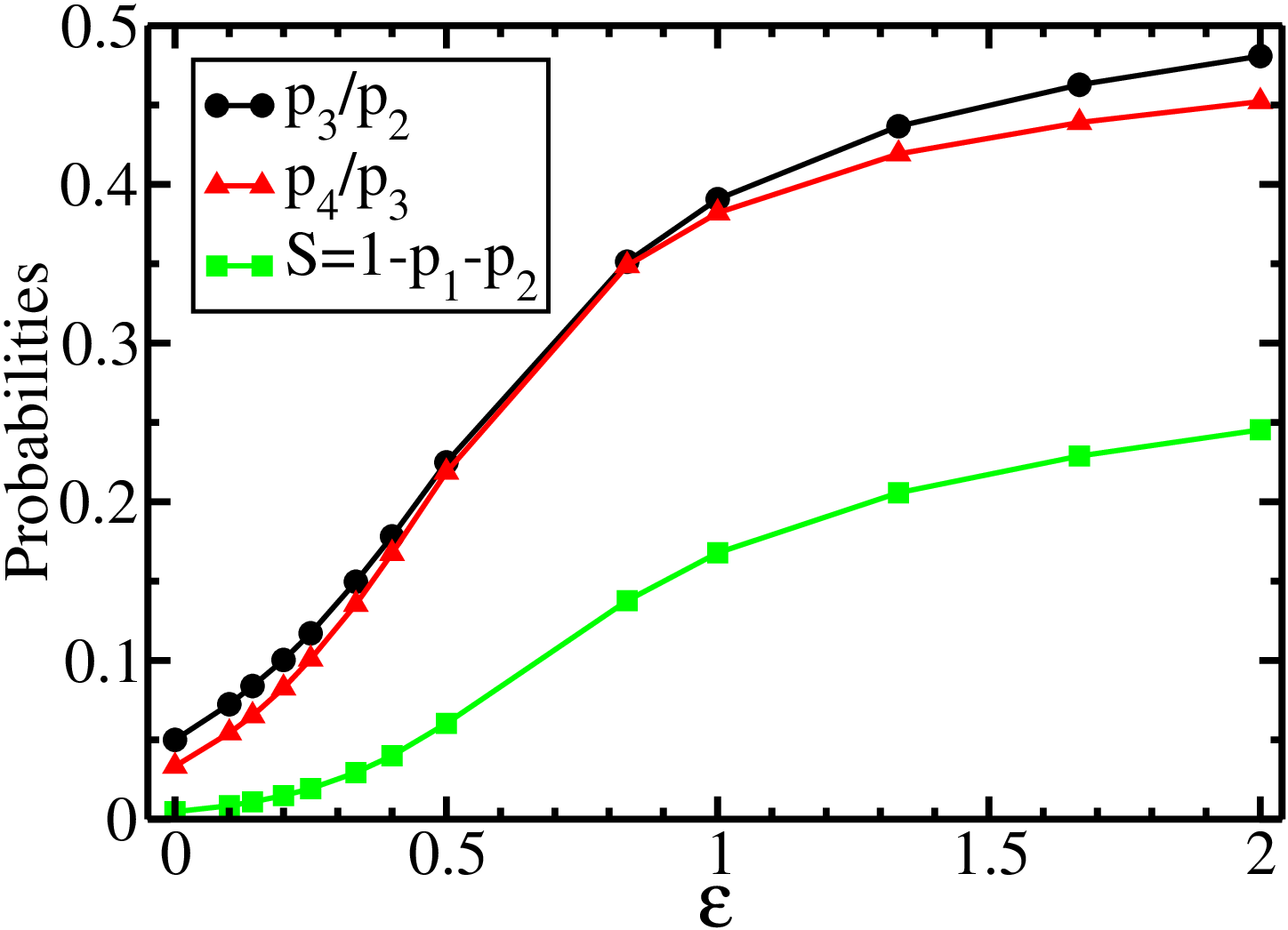}
\caption{Agent-based simulations of the probability ratios $p_3/p_2$, $p_4/p_3$, and the probability for non-binary interactions, $S=1-(p_1+p_2)$,
versus $\vre=R/\tau v_0$ at fixed density $M=0.1$ and $\eta=1.1\eta_C$.
Other parameters are the same as in Fig. \ref{fig:AGENT_vs_M}.
The values at $\vre=0$ are not from simulations but from Eqs. (\ref{PN_IDEAL}, \ref{SUM_DEF}).
}
\label{fig:AGENT_SMALL_EPS}
\end{figure}
To check this conjecture in  a more direct way, 
I perform agent-based simulations of the VM with $N=6000$ particles
near the flocking threshold in the {\em disordered phase}
and  measure the fraction of particles $p_n$ that are engaged in a $n$-particle interaction.
These fractions are time-averaged over very long runs.
For example, $p_2$ tells me the probability that a particle is part of a two-particle cluster.
Similarily, $p_3$ is the probability that a particle is interacting with exactly two others in the collision step.
These probabilities are normalized as $\sum_{n=1}^N p_n=1$.
For an ideal gas and $N\rightarrow \infty$, the probabilities are Poissonian and are given by,
\begin{equation}
\label{PN_IDEAL}
p_n^{id}={\rm e}^{-M} {M^{n-1}\over (n-1)!}
\end{equation}
In Fig. \ref{fig:AGENT_vs_M} the ratio $p_3/p_2$ as a measure of the importance of three-particle collisions is plotted
as a function of density for fixed $\Gamma=\tau v_0/R=0.03$. This value of $\Gamma$ is 
the one suggested in Vicsek's original paper
\cite{vicsek_95_97}. In addition, the quantity 
\begin{equation}
\label{SUM_DEF}
S=\sum_{n=3}^N p_n=1-(p_1+p_2)
\end{equation}
is shown as a measure of all interactions neglected by the binary collision assumption. 
It gives the fraction of all particles that are interacting with at least two others at once.
The ideal gas predictions are $p^{id}_3/p^{id}_2=M/2$ and $S^{id}=M^2+O(M^3)$.
Fig. \ref{fig:AGENT_vs_M} shows that the observed quantities are much larger than these predictions even at the lowest
density of $M=0.01$. 
This lowest density corresponds to $\rho=0.0032$ in the terminology of Refs. \cite{vicsek_95_97,nagy_07,chate_04_08}, and thus is two to three orders of magnitude smaller than typical densities used by these authors. 
However, even at this low density one sees that $p_3/p_2$ and $S$ are of order one and, therefore, more than one order 
of magnitude
larger than the ideal gas predictions.
At larger density these quantities become even larger.
While the asymptotic behavior for $M\rightarrow 0$ is not completely clear, on can still conclude that for $M\ge 0.01$
at least $42\%$ of all particles are engaged in a non-binary interaction and that the fraction of particles
involved in three-particle collisions is only a factor $\ge 0.67$ smaller than the ones undergoing binary interactions.
This means that for all practical purposes (e.g. realistic densities above $M=0.01$) the binary collision assumption is
not valid in the continuous time VM near the flocking threshold in both the ordered and disordered phases \cite{foot6}. 
As a consequence any Boltzmann approach applied to the VM close to the threshold cannot be expected to be 
quantitatively correct because it fails to correctly describe what about half of the particles do.
My results suggest that the reason for this failure is the alignment interaction that leads to a huge
increase of the collision time $\tau_0$ at small densities.

In the discrete-time VM, the divergence of the collision time for $M\rightarrow 0$, is strongly reduced by
the small ratio $\vre=R/\lambda$; 
the alignment interactions cannot keep particles together for too long. 
Instead, particles just jump away from each other after only a few microscopic
interactions.
To verify this behavior, in Fig. \ref{fig:AGENT_SMALL_EPS} the quantities $p_3/p_2$ and $S$ are shown for small $\vre$ in the disordered phase near
the flocking treshold. For $\vre\le 0.2$ only a few percent of the particles are involved in non-binary interactions.
The binary collision assumption becomes exact for $\vre \rightarrow 0$ and $M\rightarrow 0$.
Note, that the data points at $\vre=0$ were not obtained by extrapolation but by using the ideal gas predictions, Eqs. (\ref{PN_IDEAL}, \ref{SUM_DEF}). The fact, that these predictions fit perfectly into this plot with agent-based numerical data, is an additional consistency
check of the simulation. 
This plot can also be used to judge the quality of the Molecular Chaos (MC) approximation. Agreement of $S$ and $p_3/p_2$
with the ideal gas predictions, that is the data points at $\vre=0$ in Fig. \ref{fig:AGENT_SMALL_EPS}, are taken as 
indicator of the validity of MC.
For $M=0.1$ one sees that $p_3/p_2$ has already doubled at $\vre=0.2$. Thus, the
PSA approach which relies on MC but {\em not} on binary interactions, is expected to be accurate 
for $\lambda\ge 5R$. 
This is consistent with earlier results on the metric-free VM \cite{chou_12}.

To compare to the continuous-time VM,  data for large $\vre$, more specifically
for small time step $\tau$, are shown in Fig. \ref{fig:AGENT_SMALL_TAU}. %
Even if the product $\tau v_0$ is further reduced from the value $\lambda=0.03$ used in Ref. \cite{vicsek_95_97}, $p_3/p_2$ and
$S$ keep rising slowly and thus invalidate the binary collision assumption even further.
Computational limitations prevent me from investigating the limit $\tau\rightarrow 0$ in more detail.
By monitoring the global polar order parameter, 
I made sure that all data in figures \ref{fig:AGENT_vs_M}-\ref{fig:AGENT_SMALL_TAU} are taken in the disordered phase.

Since all my results are for the VM which uses point-particles
it would be interesting to see whether active matter models with more realistic interactions, for example models with additional short-range repulsion,
show a similar failure of the Boltzmann approach near the threshold to collective motion.
\begin{figure}
\centering
\includegraphics[width=6.0cm,clip]{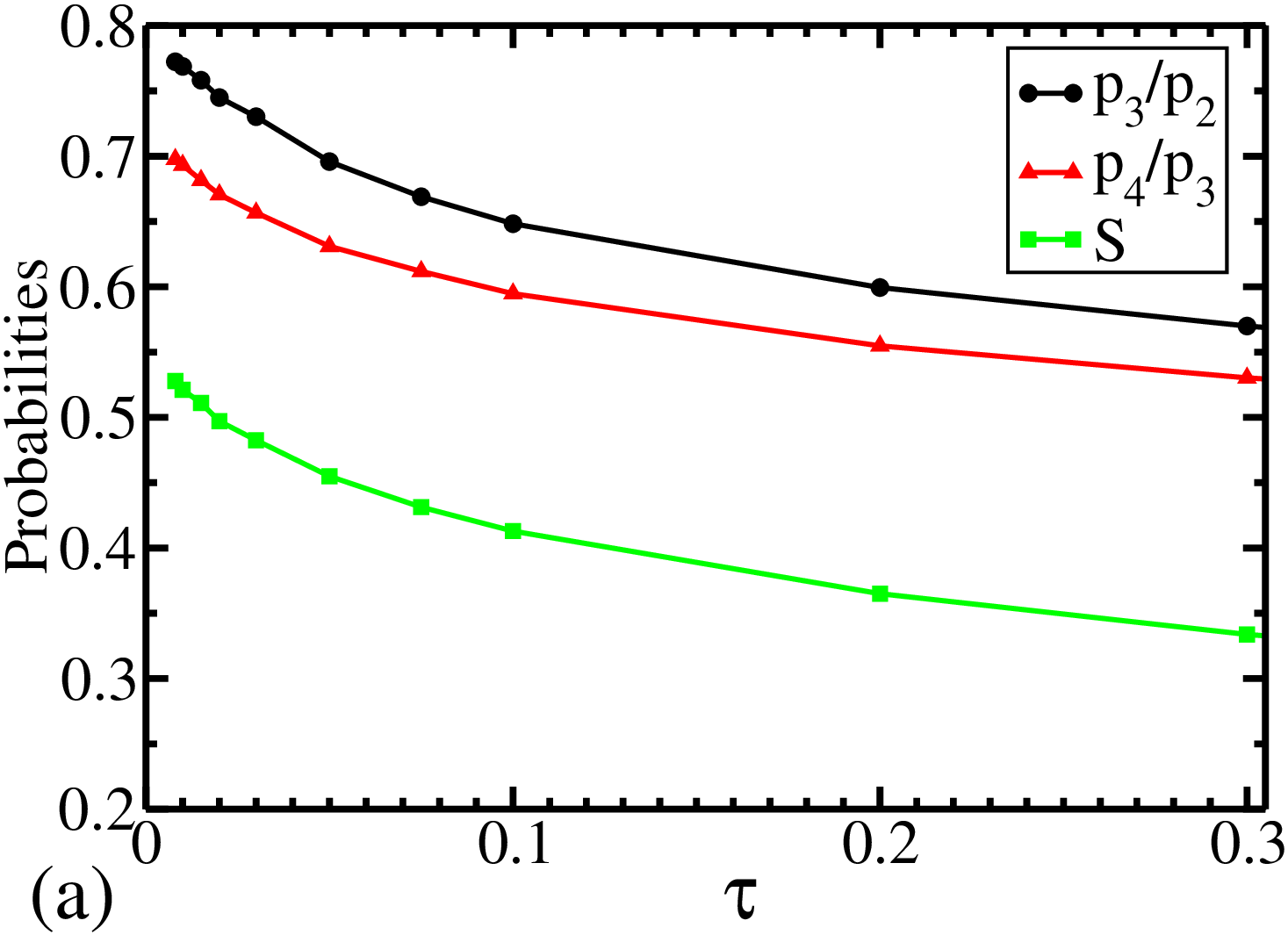}
\includegraphics[width=6.0cm,clip]{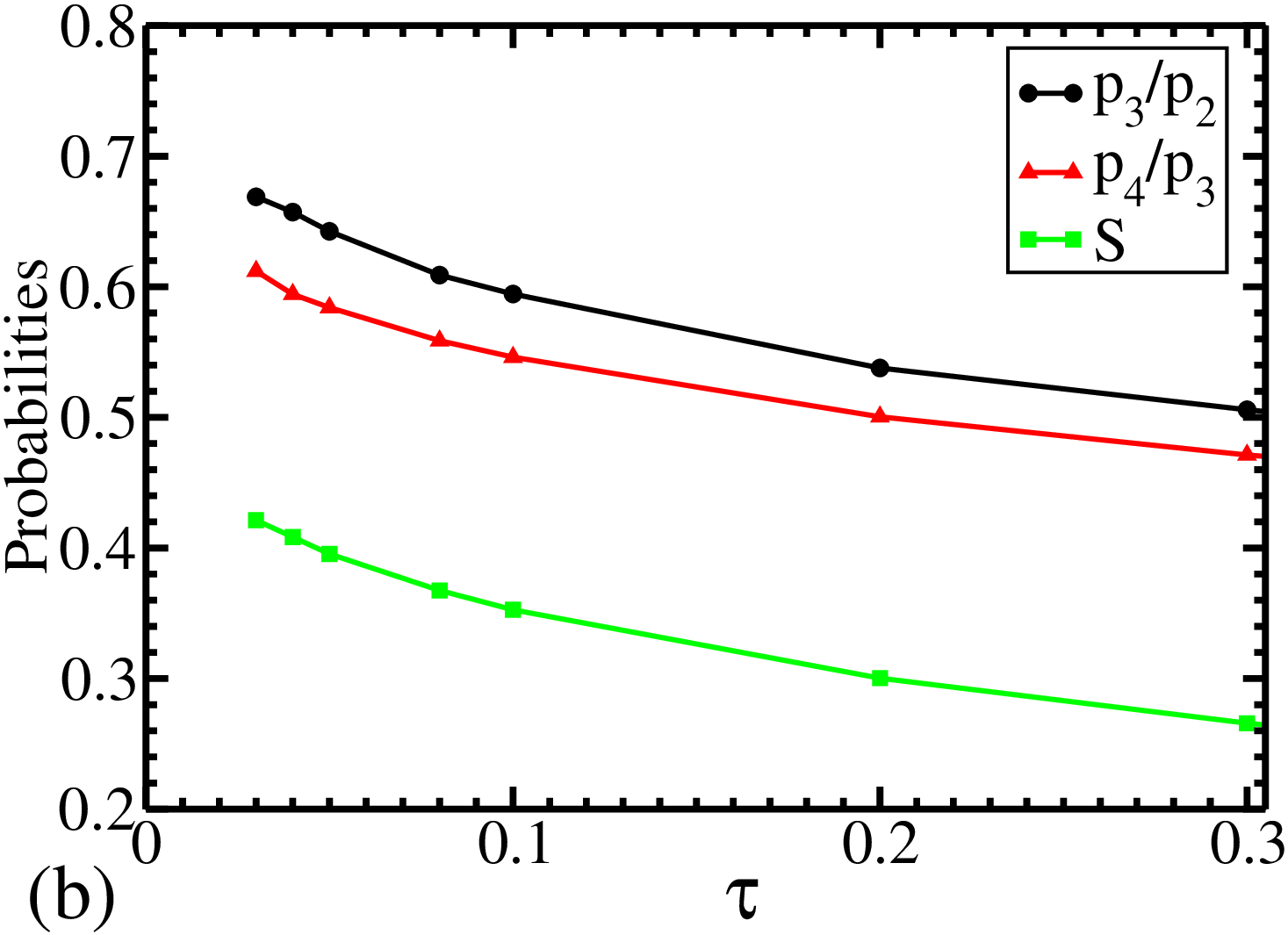}
\caption{Agent-based simulations of the probability ratios $p_3/p_2$, $p_4/p_3$, and the probability for non-binary interactions, $S=1-(p_1+p_2)$,
versus time step $\tau$ at density (a) $M=0.1$ and (b) $M=0.01$.
Parameters: $\eta=\eta_C$, for the rest of the parameters see   
Fig. \ref{fig:AGENT_vs_M}.
}
\label{fig:AGENT_SMALL_TAU}
\end{figure}

\section{Comparison to the Boltzmann model of Bertin et al.}
\label{sec:Bertin}

\subsection{Detailed mapping}

Since the Boltzmann approach by 
Bertin, Droz and Gregoire (BDG) \cite{bertin_06,bertin_09} looks similar to the kinetic equation (\ref{VLASOV}), 
it is important to understand the differences.
The BDG approach is given by
\begin{equation}
\label{BDG1}
\left({\partial\over \partial t}+{\bf v}(\theta)\cdot\nabla\right) f({\bf x},\theta,t)=I_{dif}[f]+I_{col}[f,f]
\end{equation}
featuring the convective time derivative on the left hand side.
The right hand side consists of the diffusion term
\begin{eqnarray}
\nonumber
& &I_{dif}=\lambda_d\bigg \{ -f({\bf x},\theta,t) 
+\int_{-\pi}^{\pi} d\tilde{\theta}_1 \\
&\times &\int_{-\infty}^{\infty} d\xi\,w_1(\xi)
\,\hat{\delta}[\theta-\xi-\tilde{\theta}_1]\,f({\bf x},\tilde{\theta}_1,t)\bigg \}
\label{I_DIF}
\end{eqnarray}
and the binary collision term,
\begin{eqnarray}
\nonumber
& &I_{col}=2Rv_0
\int_{-\infty}^{\infty} d\xi
\int_{-\pi}^{\pi} d\tilde{\theta}_1
\int_{-\pi}^{\pi} d\tilde{\theta}_2\;
w_2(\xi)\;
|{\bf e}(\tilde{\theta}_2)-{\bf e}(\tilde{\theta}_1)| \\
& &
\times 
f({\bf x},\tilde{\theta}_1,t)\,
f({\bf x},\tilde{\theta}_2,t)\,
\bigg\{ 
\hat{\delta}[\theta-\xi-\Phi_1]-
\hat{\delta}[\theta-\tilde{\theta}_1]
\bigg\} 
\label{I_COL}
\end{eqnarray}
where I adapted the original notation to the one used for PSA.
Obvious differences between Eqs. (\ref{BDG1}) and (\ref{VLASOV}) are that BDG is a continuous time 
approach and only considers binary collisions whereas PSA
has a discrete time step $\tau$ and can handle collisions of an arbitrary number of partners.
Assuming point particles, PSA can be applied to arbitrary density whereas a Boltzmann approach is always limited
to the limit of vanishing density.
If these were the only differences,  
in the limit of small density
one would expect 
the phase 
diagram for stationary homogeneous states to be the same. 
This is not the case \cite{ihle_11}. 

To pinpoint the fundamental difference between the models, 
let us perform the low density limit, $M\ll 1$ of Eq. ({\ref{ENSKOG1}) and neglect  
terms with $n>2$, which describe genuine interactions of three and more particles.
To ensure mass conservation, the prefactor ${\rm exp}(-M)$ is replaced by $1/(1+M)$, see supplemental material 
of Ref. \cite{ihle_13}. 
Because $f$ does not depend on position, the integral over the position of particle $2$ inside the collision circle 
can be performed.
Dividing by $\tau$ and adding $-f(\theta,t)/\tau$ on both sides, Eq. (\ref{VLASOV}) is rewritten
such that the left hand side becomes the discrete time derivative.
On the r.h.s. the following decomposition is performed
\begin{equation}
\label{SPLIT}
-{f(\theta)\over \tau}=
-{f(\theta)+Mf(\theta)\over \tau(1+M)}=
-{f(\theta)\over \tau(1+M)}-{A\over \tau (1+M)}\
\int_{-\pi}^{\pi}
d\tilde{\theta}_2\, 
f(\tilde{\theta}_2) 
f(\theta)
\end{equation}
that makes use of the equalities, 
$M=A\rho$ and $\int_{-\pi}^{\pi}d\tilde{\theta}_2 f(\tilde{\theta}_2)=\rho$.
The first term in Eq. (\ref{SPLIT}) adds a loss contibution to 
the self-diffusion term $I_1$; and the second term is incorporated into the binary collision term.
Finally, one obtains for the spatially-homogeneous PSA approach at low densities,
\begin{eqnarray}
\nonumber
& &{f(\theta,t+\tau)-f(\theta,t)\over \tau}=I_1+I_2 \\
\nonumber
& &I_1={1\over \tau(1+M)}
\bigg \{ -f(\tilde{\theta}_1,t)+
\int_{-\pi}^{\pi}
d\xi\, 
\int_{-\pi}^{\pi}
d\tilde{\theta}_1\,  
w_1(\xi)\,
f(\tilde{\theta}_1,t)
\,\hat{\delta}[\theta-\xi-\tilde{\theta}_1)]
\bigg \}
\\
\nonumber
& &I_2=
{A\over \tau(1+M)}
\int_{-\pi}^{\pi}
d\xi\, 
\int_{-\pi}^{\pi}
d\tilde{\theta}_1\,  
\int_{-\pi}^{\pi}
d\tilde{\theta}_2\,  
w_2(\xi) \\
\label{PSA_LOW}
& & \times f(\tilde{\theta}_1,t)\,
f(\tilde{\theta}_2,t)
\,
\bigg \{
\hat{\delta}[\theta-\xi-\Phi_1]-
\hat{\delta}[\theta-\tilde{\theta}_1]
\bigg \}
\end{eqnarray}
where 
$f(\theta, {\bf x}+\tau {\bf v},t+\tau)$ was replaced by $f(\theta,t+\tau)$ and
the normalization $\int_{-\pi}^{\pi}
d\xi\, w_n=1$ was used.
For stationary homogeneous states, the left hand sides of both kinetic equations vanish and we merely have to compare
the collision integrals.
The self-diffusion term $I_{d}$ of BDG becomes exactly equal
to the corresponding term $I_1$ in Eq. (\ref{PSA_LOW}) by 
choosing a self-diffusion frequency $\lambda_d=1/\tau(1+M)$.

The main difference between BDG's and PSA's collision operators is now evident:
The binary collision frequencies, that is the prefactors of the terms $I_2$ and $I_{coll}$,
do not agree. PSA's collision frequency is proportional to $1/\tau$ 
and to the area of the collision circle, $A$, 
but independent of the velocity $v_0$ and the angles of the involved particles.
The underlying physics is the one of the Vicsek model with finite time step $\tau$: particles are assumed to be 
invisible to each other during streaming and only collide once they have reached their final location.
This means when in ``flight'' they might have very close encounters
with other particles, e.g. go through each others action circles but do not interact until
the end of the streaming step. If, for example, the time step $\tau$ is reduced by a factor of ten, the particles make ten 
times more 
``stops'' during the same physical time. Hence,
the likelihood of an interaction increases by a factor of ten, and
the collision frequency increases accordingly to $w_{coll}\sim1/\tau$.

The physical picture behind the collision frequency of BDG 
is different; it 
describes the interaction rule of BDG's binary collision model:
when two particles get closer than a threshold distance, a binary interaction occurs, as outlined 
in the sentence ``In addition, binary collisions occur when the distance between two particles becomes less
than $d_0$ ..." of Ref. \cite{bertin_06}.
Since this model has a continuous time evolution, during a fixed time $T$, the focal particle engages 
in an interaction with {\em every} particle that crosses its path.
Thus, unlike in the VM, particles are never invisible to each other.
Mathematically, in the intuitive derivation 
of the Boltzmann equation, this behavior is described by a collision cylinder (or collision rectangle in 2D)
of length $v_{rel}\,\delta t$ and width $2R$ with ${\bf v}_{rel}={\bf v}_2-{\bf v}_1$ and an infinitesimal time interval $\delta t$.
Particle $2$ has to be in this collision cylinder in order to collide with the focal particle between time $t$ and $t+\delta t$.
This leads to a collision frequency proportional to $v_0$, because the faster the particles move the 
bigger is their chance 
to run into other particles during a fixed time interval.

Let's contrast this behavior with the one of the VM in the extreme limits of
vanishing and infinite particle speed $v_0$.
For $v_0=0$ and moderate to large density, there will be particles with overlapping collision circles.
Even though they cannot move, according to the rules of the VM, they will still engage in the alignment interaction.
The PSA approach has a nonzero collision frequency even for zero speed and does describe this.
However, no binary collisions occur in the BDG kinetic equation at $v_0=0$. 
This is actually common behavior for Boltzmann-like equations,
because they are all derived in the limit of vanishing density, where the likelihood of overlaps goes to zero
and no collisions can happen.
To accomodate overlaps one would have to derive higher order density corrections to the Boltzmann equation.  
In contrast, the PSA approach naturally deals with overlaps.
In the other limit of $v_0\rightarrow \infty$, the collision frequency of BDG diverges
because the focal particle runs into an infinite number of particles during its travel. 
In the VM and in PSA, $w_{coll}$ remains finite. It does not matter how many particles the focal particle passes 
during flight; it just becomes ``visible'' to others at its final location.

One could be tempted to reconcile both kinetic approaches by saying ``The BDG model might not correspond to the VM but
couldn't it just be the Boltzmann theory for a different microscopic model with continuos time dynamics?''
There is several issues with this view.

First, in the traditional derivation of the Boltzmann equation by Grad, Kirkwood, Bogolyubov from
the BBGKY-hierarchy, a coarse-graining over distances of order $R$ and times of order $\tau_0$, the 
average duration of a collision, has to be performed and the information about two-particle encounters
only enters the Boltzmann equation in a statistical statement, 
namely the scattering cross section in the collision integral
and not through the direct interaction force, or in our case, the direct alignment rule, Eq. (\ref{VM_RULE}).
In this statistical Boltzmann-sense, a collision is registered as soon as two particles enter each others
action spheres and finished if their distance is larger than the interaction range.
The coarse graining means %
that
the Boltzmann equation cannot resolve these details and only cares about in what state particles enter the sphere
and how they come out again.
This is necessary because the molecular chaos asssumption which assumes that particles are statistically independent,
can only be justified {\em before} the particles enter the action sphere. Once they are inside, during their encounter they become
increasingly correlated.
To avoid confusion, we have to distinguish between what is usually called a ``collision'' in the literature about the Vicsek-model
and a coarse-grained collision in the spirit of the Boltzmann equation. The former is just a single application of the instantaneous
aligment rule, Eq. (\ref{VM_RULE}), and will be called microscopic collision, 
whereas the latter can involve many subsequent streaming and microcopic interaction events.
In chapter \ref{sec:Escape} it was shown that the duration $\tau_0$ 
of a collision in the Boltzmann-sense can be very long in the continuous-time VM.

In the PSA approach, this 
complication does not occur
as long as the discrete time step
is much larger than $R/v_0$. Then, there is only one single microscopic collision in the 
interaction sphere and 
particles immediatly leave
the sphere. Thus, one collision in Boltzmann's definition corresponds to one collision step of the VM and
the duration $\tau_0$ is of order $\tau$.
In other words, under the condition, $\tau\gg R/v_0$, the difference between Vlasov-like and Boltzmann-like 
theories
vanishes.

However, for the continuous time VM, during one Boltzmann-collision
many of the microscopic streaming and interaction events occur.
This means, in a Boltzmann equation for this microscopic model, the scattering cross section
and not the collision kernel $w_2(\xi)\,\hat{\delta}(\theta-\Phi_1-\xi_1)$ for a single interaction should occur.

This is not how the BDG kinetic equation looks like. 
One could argue, ``Well, let's fix it, let's keep Eq. (\ref{VM_RULE}) 
as the microcopic interaction and 
let's determine the cross 
section''.
While this could be done at least numerically, the problem 
outlined in chapter \ref{sec:Escape} remains:
For all interesting applications, that is close to or inside the ordered phase, and even at very small 
densities, 
approximately half or more of all particles are engaged in non-binary interactions. 
Thus, in my opinion, there is no chance to set up a reliable Boltzmann approach for the continuous-time VM.

\newpage
\subsection{Discussion}

To summarize,

(1) I believe any Boltzmann approach for 
the Vicsek model with small discrete time step, $\tau<R/v_0$, is invalid near the flocking
threshold. This is in part due to the violation of the binary collision assumption and happens 
even in dilute systems, $0.01\le M\le 0.1$, and in the disordered phase.
Only deep in the disordered phase, far away from the threshold, a valid Boltzmann equation could be set up
but that is not interesting. Here, the term ``Boltzmann approach'' refers to a description solely based on the 
one-particle distribution
function $f(\theta,{\bf x},t)$, which only considers spatially-local single and binary collisions. 

(2) The Boltzmann-inspired approach of BDG \cite{bertin_06} does neither correctly 
describe the continuous-time nor the discrete-time
Vicsek model on the quantitative level because it seems to
inconsistently mix a Boltzmann-like collision frequency with a Vlasov-like interaction kernel.
However, on the plus side, it is easier to handle than the PSA approach and has delivered 
important qualitative results \cite{bertin_06,ngo_12,peshkov_12}.

Adamant users of BDG might justify the interaction kernel $w_2(\xi)\,\hat{\delta}[\theta-\xi-\Phi_1]$ 
as
an already averaged
mesoscopic cross section. However, this just leads to more questions such as, 
(i) what is the underlying microscopic interaction leading to this 
cross section, and is this interaction consistent with the physics of any real binary collision, and 
(ii) would this interaction violate the validity of the Boltzmann approach?
My guess is, that even if one can reverse-engineer the underlying microscopic rule, 
the same will happen that was found in chapter \ref{sec:Escape}: near the flocking threshold the collisions (defined in the Boltzmann-spirit)
will take too
long, thus leading to clusters with three and more particles even if the overal density is low.

One might wonder why Ref. \cite{bertin_09} reported decent agreement between the BDG theory and 
agent-based simulations of
the discrete time VM.
I believe this was coincidence 
because for fixed $R$ and $\lambda_d$ there is one value of the velocity
$v_0$ where the collision integrals of both models approximately agree.
The condition is $\pi R^2=2R\langle|{\bf v}_1-{\bf v}_2|\rangle/\lambda_d\approx 2R\sqrt{2}v_0/\lambda_d$.
If this is fulfilled, the ratios $I_1/I_2$ and $I_d/I_{col}$ that determine the phase diagram
of stationary, homogeneous solutions  at low density, are approximately the same.
As shown in Fig. 1 of Ref. \cite{ihle_11} large discrepancies can occur if one chooses other velocities
or higher densities.

Another question that comes to mind is: if one were to send the time step $\tau$ 
to an infinitesimal value in the VM,
one would recover the continuous time VM, and, formally, one could also write down the PSA approach
for such a small time step. 
The two theories would then attempt to describe the same system, 
how come they still look different?
The first part of the answer is: 
the PSA approach simply ceases to be valid if the time step violates 
the condition $v_0\tau/R\gg 1$. 
Second, the BDG kinetic equation is inconsistent with 
the microscopic collision rules, Eq. (\ref{VM_RULE}), of the VM.
This is because a Boltzmann-equation in the traditional sense 
contains statistical information in form of 
a scattering cross section and not the scattering kernel of a single alignment interaction.
This scattering cross section is the result of many streaming and alignment steps,
that take place during the collision time interval $\tau_0$, see chapter \ref{sec:Escape}.
Third, the PSA approach resolves the microscopic time scale like the Vlasov-equation, 
whereas a Boltzmann approach  
works on a coarse-grained manifold, and therefore should look different. 

While the PSA approach delivers quantitative agreement for $v_0\tau/R=1/\vre\gg 1$ and arbitrary density,
one might still be tempted to dismiss it because of its unphysical feature that particles can ``tunnel'' through each other
during streaming.
I see this is as the price one has to pay to obtain a kinetic theory that is valid near the flocking threshold.
Furthermore, this approach is just the zeroth order contribution 
in the expansion parameter $\varepsilon=R/v_0\tau$ of a more general theory for the VM. 
The next correction 
in $\varepsilon$, which contains clustering effects and goes beyond Molecular Chaos,
will be presented elsewhere, \cite{chou_14}.  

\section{Summary}

In this discussion \& debate paper, 
a recent kinetic approach for active particles is reviewed. 
For simplicity, I focus on the Vicsek-model (VM) as
a paradigm of active matter.
The kinetic theory approach is named ``Phase Space Approach'' (PSA) because it is
based on an exact Chapman-Kolmogorov equation in phase space. 
It is designed to handle
discrete time dynamics and multi-particle interactions
that are given by collision rules and are not required
to follow
from a Hamiltonian.
The discrete time step of the VM-algorithm is utilized 
to turn the molecular chaos assumption into a controlled and tunable
approximation.
This approximation is used to obtain a nonlocal 
mean-field theory for the one-particle distribution function.

Hydrodynamic equations for the PSA approach are derived by means of a third-order Chapman-Enskog expansion
using a non-traditional scaling of the temporal derivatives.
The equations are compared to the Toner-Tu theory of polar active matter. New terms, that seem to 
be absent in the Toner-Tu theory, are emphasized.
Common convergence problems of Chapman-Enskog and similar gradient expansions are pointed out and
possible remedies are discussed.

The average duration $\tau_0$ of a collision of two particles that follow Vicsek's alignment rule
is measured in Monte Carlo simulations.
It is found that this time scales with nearly the same power of the density than
the mean free time between collisions, $\tau_{coll}$. Thus, if density is decreased, 
the ratio $\tau_0/\tau_{coll}$ does not
go quickly to zero as in regular gases. This suggests that the binary collision approximation -- a
key ingredient of a Boltzmann approach -- is 
not even valid in dilute systems
because
collisions take so long that typically
other particles join ongoing binary encounters.

This hypothesis is confirmed by
agent-based simulations
of the standard VM. In these simulations, the fraction of particles that are engaged in non-binary 
interactions is recorded and turns out to be quite large.
Therefore, Boltzmann approaches are not suitable for quantitative descriptions of the continuous-time VM near the transition to collective motion.

The Boltzmann approach of Bertin {\em et al.} (BDG), \cite{bertin_06,bertin_09} 
is critically assessed and compared term-by-term to the PSA approach.
I find that even at small densities and in homogeneous systems there is a significant 
difference between
PSA and BDG:
the collision frequencies of the binary collision terms depend on different physical parameters. 
I present arguments to substantiate my opinion that the 
approach of Refs. \cite{bertin_06,bertin_09} is not a consistent description of a VM-like microscopic model and thus
not able
to produce quantitatively correct results for the VM in any limit.
I also argue that it is not worth to make it consistent because of the 
general problems of Boltzmann approaches for systems with alignment interactions.

Support
from the National Science Foundation under grant No.
DMR-0706017 
is gratefully acknowledged.
Computer access from the North Dakota State University Center for Computationally Assisted Science and Technology and the Department of Energy through Grant No. DE-FG52-08NA28921, and Grant No. DE-SC0001717 is 
gratefully acknowledged.
I would like to thank Henk van Beijeren and Yen-Liang Chou for valuable discussions.


\begin{thebibliography}{99}
\bibitem{foot3}
In principle, $f$ could have, and probably in reality does have, 
a true nonlocal dependence on the hydrodynamic fields.
In practice, $f$ is expanded in spatial gradients of these fields and the expansion is truncated 
somewhere,
forcing the dependence to be local.
\bibitem{foot5}
For example, the pronounced finite size effects of the transition to collective motion, including
the strong system size dependence of steep solitons that show hysteresis and lead to a discontinuous flocking transition, see Ref. \cite{ihle_13},
have 
not yet been reproduced by this theory.
\bibitem{foot6}
Additional simulations in the ordered phase show that the likelihood for non-binary interactions 
becomes even larger compared 
to the disordered phase with similar parameters.
\bibitem{foot4}
Typical parameters in Ref. \cite{vicsek_95_97} are $v_0=0.03$, $\tau=R=1$, $\rho_0=0.4$
leading to $l_D=1.58$, $\lambda=0.03$, and $M=1.257$.
\bibitem{foot1}
$P$ refers to an ensemble of independent Vicsek systems which are initialized
at time $t=0$ with some inital probability density $P_0$. 
This inital density is assumed to be symmetric
aginst permuting particle indices.
\end{thebibliography}
\end{document}